\documentclass[pdflatex,sn-mathphys-num]{sn-jnl}

\usepackage{graphicx}%
\usepackage{multirow}%
\usepackage{amsmath,amssymb,amsfonts}%
\usepackage{amsthm}%
\usepackage{mathrsfs}%
\usepackage[title]{appendix}%
\usepackage{xcolor}%
\usepackage{textcomp}%
\usepackage{manyfoot}%
\usepackage{booktabs}%
\usepackage{algorithm}%
\usepackage{algorithmicx}%
\usepackage{algpseudocode}%
\usepackage{listings}%
\usepackage{natbib}
\usepackage{enumerate}
\usepackage{caption}
\usepackage{float}
\usepackage[utf8]{inputenc}

\theoremstyle{thmstyleone}%
\theoremstyle{thmstyletwo}%

\theoremstyle{thmstylethree}%

\begin{document}

\title[Distinguishing life from non-life via molecular frontier orbital energy gaps]{Distinguishing life from non-life via molecular frontier orbital energy gaps}


\author[1]{\fnm{José L.} \sur{Ramírez-Colón}}\email{jcol6@gatech.edu}
\equalcont{These authors contributed equally to this work.}

\author[2]{\fnm{Ziqin} \sur{Ni}}\email{zni47@gatech.edu}
\equalcont{These authors contributed equally to this work.}

\author*[1,2]{\fnm{Christopher E.} \sur{Carr}}\email{cecarr@gatech.edu}
\equalcont{These authors contributed equally to this work.}

\affil[1]{\orgdiv{School of Earth and Atmospheric Sciences}, \orgname{Georgia Institute of Technology}, \orgaddress{\street{311 Ferst Dr}, \city{Atlanta}, \postcode{30332}, \state{GA}, \country{USA}}}

\affil[2]{\orgdiv{School of Aerospace Engineering}, \orgname{Georgia Institute of Technology}, \orgaddress{\street{270 Ferst Dr}, \city{Atlanta}, \postcode{30332}, \state{GA}, \country{USA}}}


\abstract{
Amino acids (AAs) are a key target in the search for life beyond Earth \cite{neveu_ladder_2018, davila_chance_2014, cieslarova_microorganisms_2022} due to their extensive role in the machinery of all known life, persistence over geologic timescales \cite{glavin_abundant_2025,parker_extraterrestrial_2023}, and analytical detectability \cite{glavin_abundant_2025, ohshiro_detection_2014, benhabib_multichannel_2010, ligterink_origin_2020}. However, AAs can also arise from abiotic processes on planets and in space  \cite{glavin_abundant_2025, parker_extraterrestrial_2023}. For example, material from asteroid Bennu contained 33 AAs, including 15 of the 20 proteinogenic AAs that are fundamental to life's functions \cite{glavin_abundant_2025, mojarro_prebiotic_2025}. Distinguishing life from non-life based on AAs in a sample remains an unsolved problem, particularly when their isotopic and structural signatures (e.g., chirality) could be altered via physicochemical processes \cite{steel_abiotic_2017}. Here we introduce LUMOS (Life Unveiled via Molecular Orbital Signatures), a statistical framework that distinguishes life from non-life by analyzing the distribution of abundance-weighted HOMO-LUMO gap (HLG) values of AAs within a sample. Compilation of AAs datasets from diverse environments and provenances revealed that abiotic samples display highly uniform distributions of AAs HLGs. In contrast, biotic samples show greater variance and preference towards AAs with lower HLG, likely reflecting the need for life to control when, where, and how chemical reactions occur. LUMOS achieves $>$95\% accuracy in distinguishing biotic versus abiotic provenance across diverse environmental and extraterrestrial conditions. These results suggest that varied molecular reactivity within biochemical systems may be a universal feature of life, representing an agnostic biosignature unlinked to the specific set of AAs used by life as we know it. LUMOS is compatible with existing analytical instrumentation, applicable to returned samples or \textit{in situ} analyses. Broader characterization of abiotic and biotic environments will further refine the chemical boundaries separating biotic from abiotic chemical systems.}




\maketitle

\section{Introduction}\label{Introduction} 

Chemical reactivity emerges from the quantum behavior of electrons, which occupy discrete orbitals around atomic nuclei and reorganize when atoms form molecules \cite{mulliken_electronic_1932}. This reorganization creates new molecular orbitals that govern how electrons are shared or transferred during chemical reactions \cite{pauling_nature_1932}. Among these, the highest occupied and lowest unoccupied molecular orbitals (HOMO and LUMO), also referred to as frontier orbitals, were recognized in the mid-20th century as dominant predictors of chemical reactions \cite{fukui_molecular_1952,fukui_role_1982, hoffmann_selection_1965}. The energy gap between them, the HOMO-LUMO gap (HLG), reflects how readily a molecule can exchange or rearrange electrons: small gaps favor instability, while large gaps indicate electronic stability \cite{fukui_role_1982}. Because these frontier orbitals shape how molecules interact, they strongly influence reaction pathways and the stereochemistry of products \cite{nakliang_emerging_2021}, providing a general quantitative framework for comparing reactivity across chemical systems.

Life is commonly defined as a self-sustaining chemical system capable of Darwinian evolution \cite{benner_2010,vitas_towards_2019}. Such a system must regulate when, where, and how molecular interactions occur. Because this regulation depends on controlling reactivity, a broad distribution of molecular HLGs is likely a universal hallmark of any biotic chemical system. Amino acids (AAs) have long been considered potential biosignatures \cite{neveu_ladder_2018,davila_chance_2014,cieslarova_microorganisms_2022,georgiou_2018}, yet their ability to arise from both abiotic \cite{herbst_complex_2009,parker_extraterrestrial_2023, aponte_pahs_2023, glavin_abundant_2025, miller_production_1953} and biotic pathways poses a fundamental challenge to distinguishing their provenance (Supplementary Background \ref{ref:suppbackorganics}). Existing approaches to distinguishing life from non-life, ranging from isotopic analyses \cite{mccollom_carbon_2006,glavin_unusual_2012} to abundance patterns \cite{cieslarova_microorganisms_2022,cleaves_robust_2023} and  molecular complexity metrics \cite{marshall_probabilistic_2017,marshall_identifying_2021}, are often constrained by overlapping distributions, degradation effects \cite{truong_decomposition_2019, steel_abiotic_2017,johnson_metal-catalyzed_2010}, and methodological limitations, highlighting the need for more robust and universally applicable biosignature frameworks (Supplementary Background \ref{ref:suppbackapproaches}). Here we show that AAs HLGs constitute a quantitative biosignature, with abundance-weighted HLGs distributions reliably distinguishing samples of biotic and abiotic provenance, directly addressing the chemical system aspect of life.

\section{Results}\label{Results} 

\subsection{The HOMO-LUMO gap distribution is a biosignature}\label{results:HLGbiosignature}
\begin{figure}[H]
    \centering
    \includegraphics[width=1\textwidth]{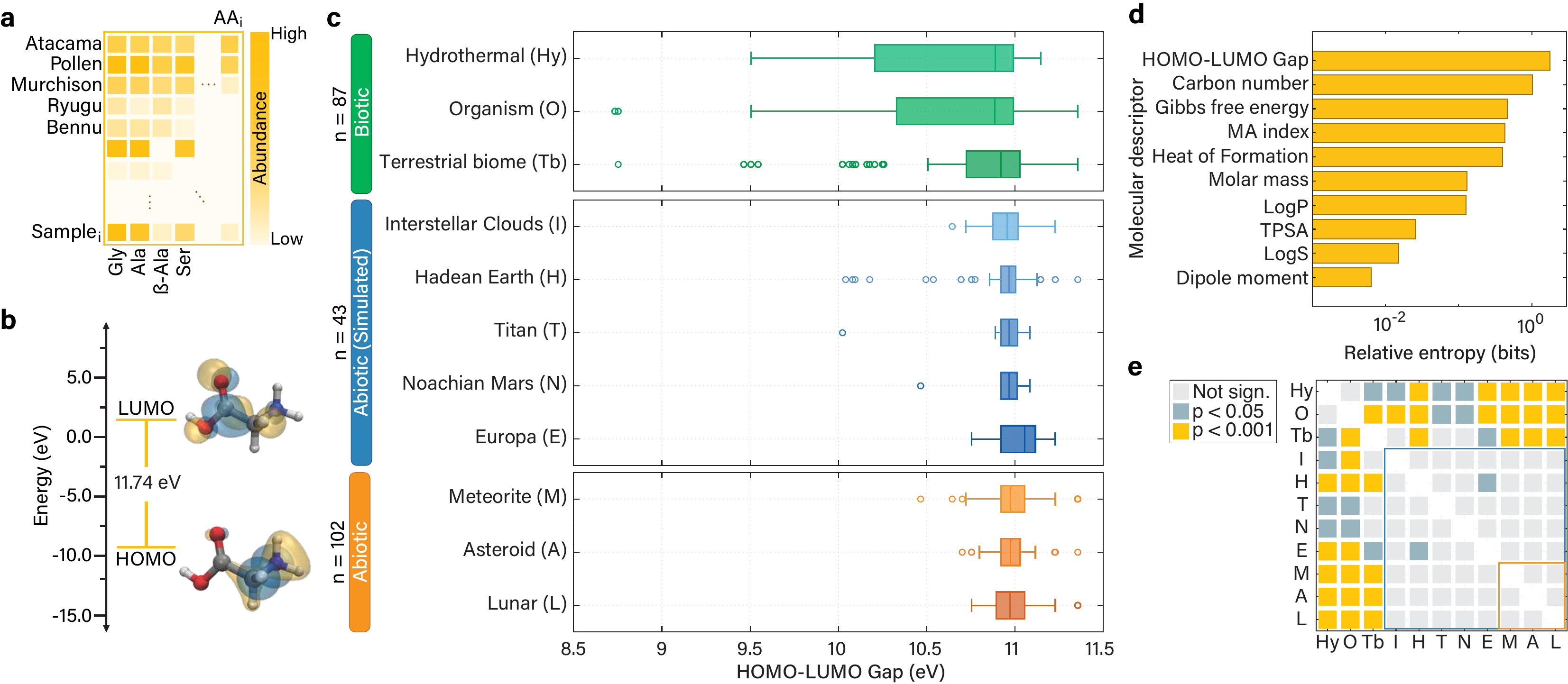}
    \caption{\textbf{Distribution of HOMO–LUMO gaps in biotic and abiotic amino acids environments.} \textbf{a}, Construction of an amino acid database from 102 abiotic and 87 biotic samples with abundance measurements, plus 43 simulated abiotic samples without abundance measurements. \textbf{b}, Quantum chemical calculations to determine the HLG for each amino acid. As an example, the frontier molecular orbitals of glycine are illustrated. The blue and yellow lobe colors correspond to the positive and negative phases of the wavefunction, respectively. The orbitals are visualized with VMD\cite{HUMP96} using results from $\omega$B97XD/def2-TZVP/PCM(Water) calculations performed with Psi4\cite{smith_psi4_2020}. \textbf{c}, Distribution of HLG values [as determined by the $\omega$B97XD/def2-TZVP/PCM(Water) method] for amino acids across three categories: abiotic, abiotic-simulated, and biotic, with corresponding boxplots displaying the statistical distributions for each subcategory. \textbf{d}, Symmetric relative entropy of selected molecular descriptors highlighting separation ability in distinguishing between biotic and combined abiotic categories. \textbf{e}, Statistical significance matrix showing pairwise comparisons among the subcategories.
}
    \label{fig:HLGdistribution}
\end{figure}

Biotic and abiotic samples have AAs with distinct diversity, range, and magnitude of HLG values (Fig. \ref{fig:HLGdistribution}). We compiled a database (Fig. \ref{fig:HLGdistribution}a) of amino acid profiles from 87 environmental samples (``biotic''), 102 abiotic samples (``abiotic''), and 43 abiotic experimental simulations (``abiotic simulated'') (Methods \ref{methods:database}). We estimated the HLG of each of the 64 AAs represented within the database (Fig. \ref{fig:HLGdistribution}b) using a computationally-intensive reference density functional theory (DFT) method ($\omega$B97XD/def2-TZVP) and a fast semi-empirical method (MNDO) (Methods \ref{methods:qc}). We then analyzed the distribution of AAs HLGs across the three categories. Biotic samples have a smaller set of AAs, with only 37 unique identifications, but the range of HLG values is nearly double (max-min, 2.622 eV) relative to those from abiotic and abiotic-simulated categories (1.315 eV) (Fig. \ref{fig:HLGdistribution}c). AAs with HLGs below 10 eV are exclusive to biotic samples (lowest value of 8.73 eV), whereas the threshold HLGs of AAs in abiotic and abiotic-simulated samples are 10.46 eV and 10.02 eV, respectively. Compared to other molecular descriptors that we examined (Fig. \ref{fig:HLGdistribution}d, Methods \ref{methods:descriptors}), HLG stands out as the most effective property in distinguishing biotic versus abiotic samples with a symmetric relative entropy of 2.61 bits.

The distinct HLG distributions between biotic and abiotic samples were recapitulated when examining HLG distributions according to specific sampling locations or environments, denoted subcategories (Fig. \ref{fig:HLGdistribution}c). Consistent patterns were observed independent of sample quantity or amino acid occurrence frequency (Supplemental Dataset 1, Statistics\_Tables, Table 4). All biotic subcategories had HLG range $>$1.5 eV and standard deviation $>$0.4 eV, compared to the range $<$1.3 eV and standard deviation $<$0.2 eV of all abiotic subcategories. All abiotic subcategories show no pairwise statistically significant difference with the exception of Hadean Earth and Europa. Most biotic subcategories had significant differences from abiotic subcategories (Fig. \ref{fig:HLGdistribution}e, $p<0.001-0.05$ for 21 of 24 pairwise comparisons), suggesting distinct amino acid HLG patterns between biotic and abiotic samples.

\subsection{Abundance-weighted metrics optimize class separation}\label{subsec2}
\begin{figure}[H]
    \centering
    \includegraphics[width=1\textwidth]{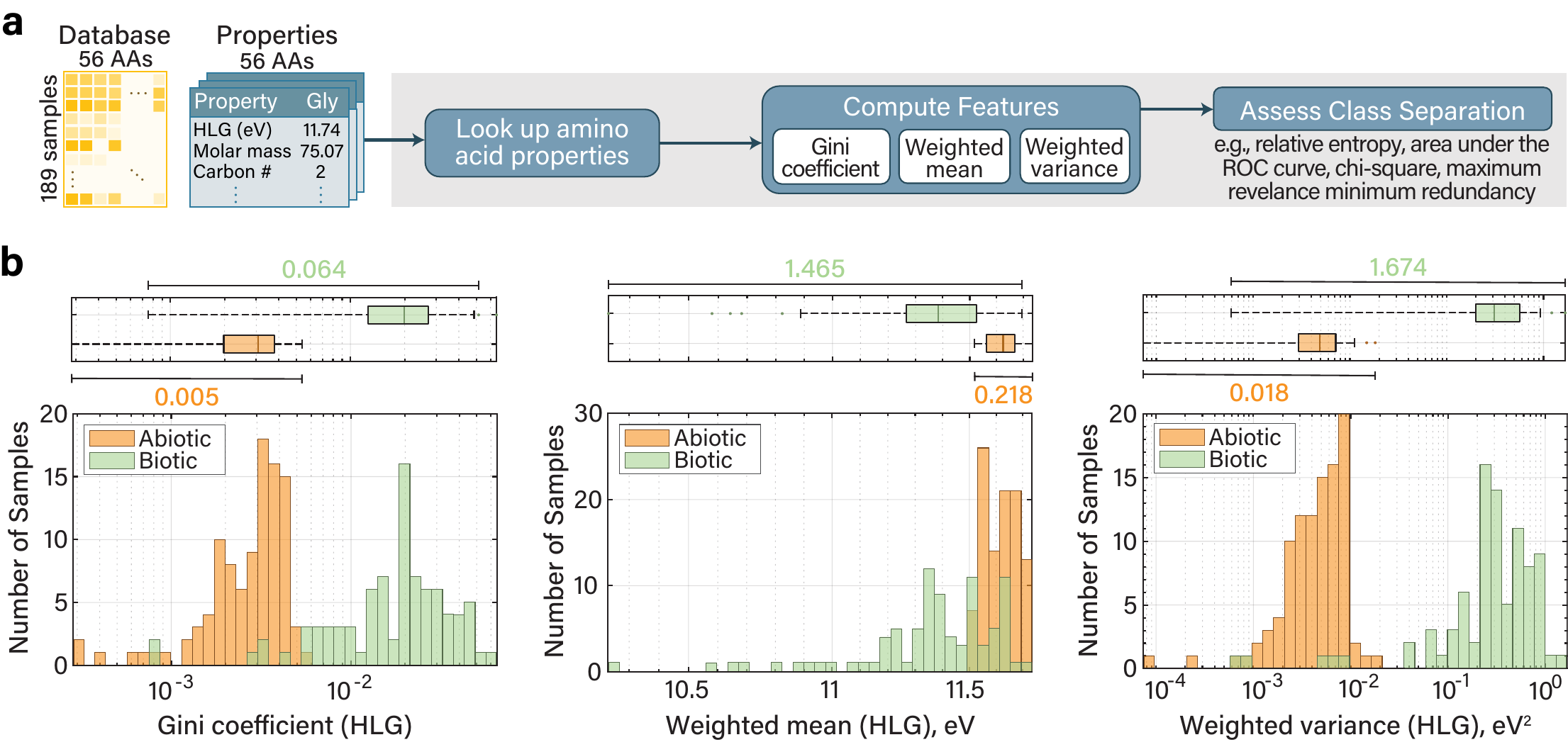}
    \caption{\textbf{Abundance-weighted molecular descriptors improve classification of biotic and abiotic amino acid samples.} \textbf{a}, Workflow for assessing the separation between biotic and abiotic samples using a database of amino acid abundances and a molecular properties dataset spanning 56 amino acids. Each amino acid in a sample is matched to a molecular descriptor of interest, which is then weighted by its abundance using one of three approaches: Gini coefficient, weighted mean, or weighted variance. The resulting values are used to evaluate the ability of the descriptor to distinguish between sample classes. \textbf{b}, Distribution of Gini coefficient (log scale), weighted mean, and weighted variance (log scale) values across samples using the HLG. Box plots above each distribution indicate the spread and central tendency of values across biotic and abiotic samples.
}
    \label{fig:abundanceweighted}
\end{figure}

Weighting molecular descriptors by their abundance dramatically improves the separation between biotic and abiotic categories (Fig. \ref{fig:abundanceweighted}). While AA abundance alone showed increased overlap between biotic and abiotic samples (Supplemnetary Fig. \ref{fig:TotalAbundance}a), incorporating molecular descriptor information significantly enhanced classification performance. We computed the Gini coefficient, weighted mean, and weighted variance for each descriptor, resulting in three features per descriptor (Fig. \ref{fig:abundanceweighted}a, Methods \ref{methods:separation}). For the HLG, the biotic category exhibits ranges that are both shifted and 12, 6, and 92 times larger than those of the abiotic category, respectively (Fig. \ref{fig:abundanceweighted}b). As a result, stronger class separation is achieved (symmetric relative entropy of 250.38, 25.362, and 19,232 bits, respectively). This weighting captures the proposed prior use of the AA abundance distribution as a biosignature (e.g., \cite{davila_chance_2014,cieslarova_microorganisms_2022}).

\subsection{HOMO-LUMO gap metrics outperform others}\label{results:HLGoutperform}

\begin{figure}[H]
    \centering
    \includegraphics[width=1\textwidth]{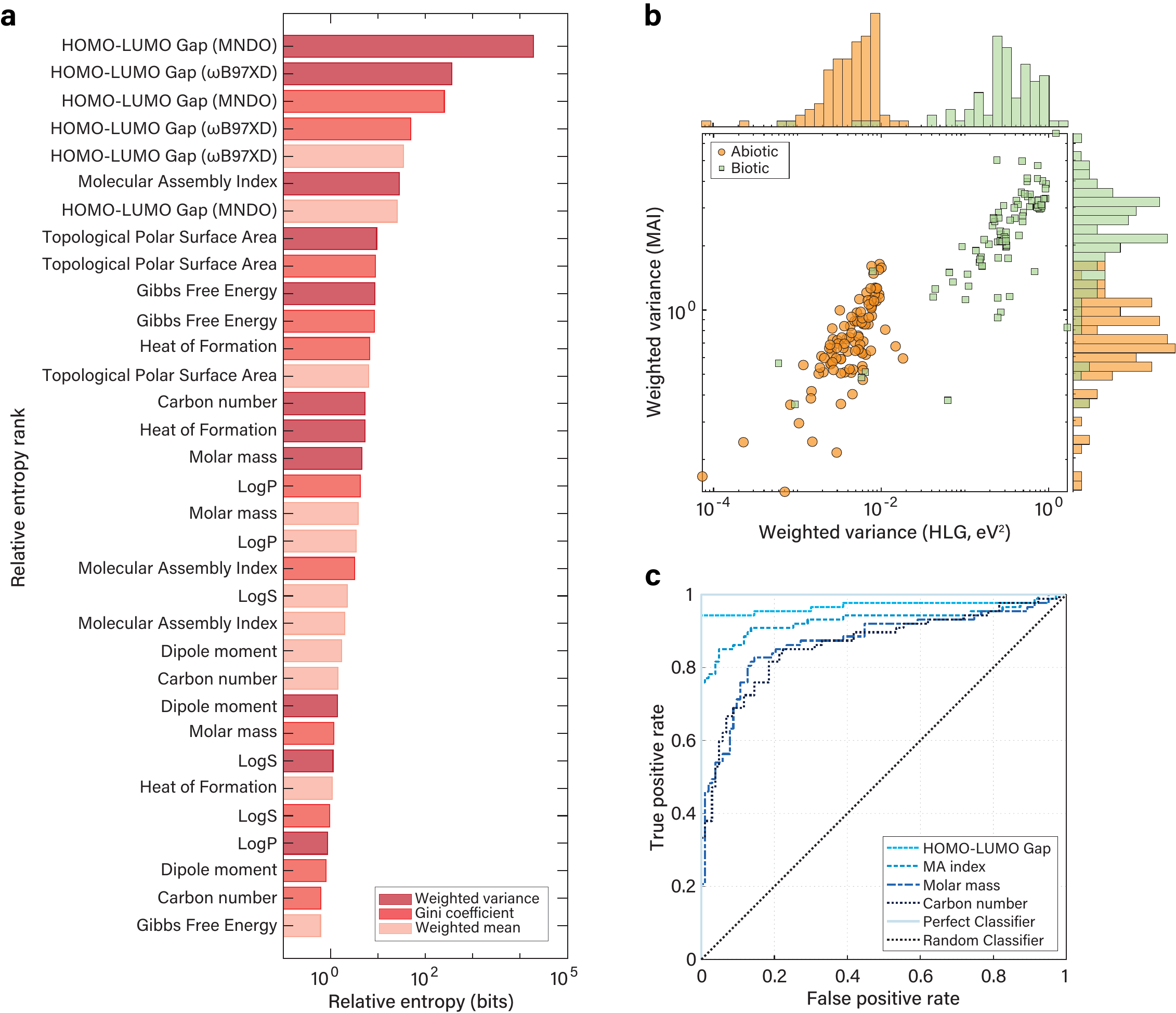}
    \caption{\textbf{Performance of abundance-weighted molecular descriptors in biotic–abiotic classification}. \textbf{a}, Descriptor performance ranked through symmetric relative entropy analysis of statistical metrics (weighted mean, weighted variance, Gini coefficient) applied to abundance-weighted descriptors. \textbf{b}, Two-dimensional scatter plot of weighted variance for HLG (MNDO method) versus Molecular Assembly (MA) Index (next-best non-HLG descriptor) across biotic and abiotic samples, with histogram displaying the distribution of the weighted variance (wvar) values of each samples across both molecular descriptors. \textbf{c}, Receiver operator characteristic (ROC) curves comparing classification performance of wvar for HLG (AUC=0.968), MAI (AUC=0.930), molar mass (AUC=0.869), and carbon number (AUC=0.863).}
    \label{fig:performance}
\end{figure}

The HLG achieved the strongest class separation among all evaluated molecular descriptors using abundance-weighted metrics as quantified by symmetric relative entropy (Fig. \ref{fig:performance}a; Supplemental Dataset 1, Statistics\_Tables, Table 5). The superior performance of HLG was maintained for both quantum chemical computational methods: MNDO and $\omega$B97XD/def2-TZVP. Features based on the electronic property of HLG consistently outperformed surface, energetic, physical, and solubility-based properties (Supplementary Fig. \ref{fig:ClassificationMethods}a). This finding was further validated using multiple class separation methods, including area under the Receiver Operator Characteristic (ROC) curve (Supplementary Fig. \ref{fig:ClassificationMethods}b), chi-square (Supplementary Fig. \ref{fig:ClassificationMethods}c), and Maximum Relevance Minimum Redundancy (Supplementary Fig. \ref{fig:ClassificationMethods}d). Among the three metrics, weighted variance generally provided the greatest separation across molecular descriptors (Fig. \ref{fig:abundanceweighted}b, Fig. \ref{fig:performance}a, Supplemental Dataset 1, Statistics\_Tables, Tables 7 and 8). Exceptions included heat of formation, LogP, LogS, and dipole moment, where the Gini coefficient or weighted mean were more effective.

The weighted variance of the HLG separates biotic and abiotic samples, with only $2.72\%$ (5 samples, Jaccard Index) overlap (Fig. \ref{fig:performance}b). Subsequent analysis revealed that the overlapping samples are anomalies, suggesting that overlap may be further reduced with appropriate filtering criteria (Supplemental Discussion, Section \ref{supdis:wvar}). In contrast, the next highest performing feature, the molecular assembly index (MAI) weighted variance, has four times more overlap (19 samples, 11.1\%). Incorporating the MAI provides no additional separation (Fig. \ref{fig:performance}b). Receiver operator characteristic (ROC) area under the curve (AUC) analysis further confirmed HLG's separation capability, outperforming MAI by 4\%, and molar mass and carbon number by 11\% (Fig. \ref{fig:performance}c). The HLG metric maintained a true positive rate of 0.942 at zero false positive rate, highlighting its high sensitivity and specificity, respectively. Machine learning models using only the HLG-MNDO weighted variance were able to achieve high accuracy ($96.8\pm2.4\%$, coarse tree model) in classification, with no additional accuracy achieved when including all 33 features for this representative interpretable model (Methods \ref{methods:ml}, Supplementary Table~\ref{table:testaccuracies}).

\subsection{LUMOS assesses confidence in biogenicity}\label{LUMOS}

\begin{figure}[H]
    \centering
    \includegraphics[width=1\textwidth]{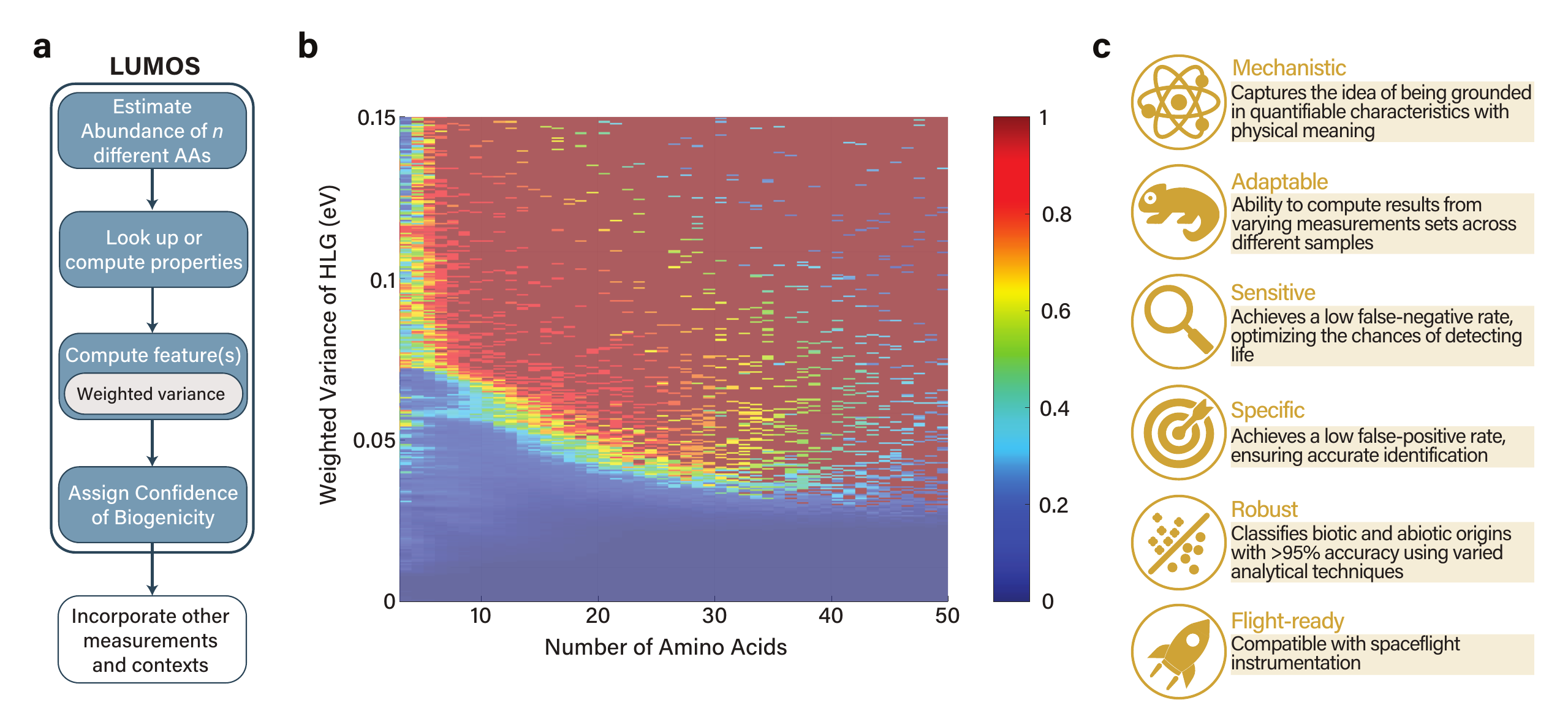}
    \caption{\textbf{Overview and performance of the LUMOS framework for biogenicity assessment.} \textbf{a}, The LUMOS (Life Unveiled using Molecular Orbital Signatures) framework integrates amino acid abundance measurements with quantum chemical descriptors to evaluate biogenicity. First, the abundance of a set of amino acids is estimated, followed by the computation of molecular descriptors (e.g., HOMO–LUMO gap). These descriptors are then weighted by abundance using statistical features such as weighted variance. The resulting value is used to assess the confidence in biogenicity, informed by additional contextual data and environmental provenance. \textbf{b}, Heatmap of  $P(B|E)$, the confidence that a sample is biotic ($B$) given the observed weighted variance (evidence $E$) as a function of the number of amino acids measured. The prior $P(B)$ is taken as $0.01$. Color intensity reflects the biogenicity probability as determined by Bayesian inference applied to simulated distributions of biotic and abiotic samples. \textbf{c}, Criteria met by the LUMOS framework for distinguishing biotic from abiotic amino acid systems.}
    \label{fig:LUMOSframework}
\end{figure}

We created LUMOS (Life Unveiled via Molecular Orbitals Signatures) to determine the likelihood of unknown samples to be biological based on their abundance-weighted HLG values (Fig. \ref{fig:LUMOSframework}a), thereby establishing a guideline for applying the HLG measurements in life detection missions. LUMOS uses Bayesian analysis to calculate confidence scores for biotic origin, an approach we implemented by running extensive  Monte Carlo simulations (n = 1,000,000) on our AAs dataset (Methods \ref{methods:sml}). We simulated samples containing 3 to 50 AA (\textit{n}) to evaluate how sample size affects biosignature detection. We focused on the HLG (MNDO) features, without loss of generality, because the other features analyzed were found to be redundant (Supplementary Table~\ref{table:testaccuracies}).

The simulations revealed clear differences between biotic and abiotic systems across all three HLG-based metrics (Supplementary Fig. \ref{fig:PofE}). Simulated abiotic samples remained below Gini coefficient values of 0.005 (99\% across all \textit{n} values, Supplementary Fig. \ref{fig:PofE}a), while biotic samples consistently exhibited higher Gini coefficients ($99\% > 0.01$ for \textit{n} $\geq 20$,  Supplementary Fig. \ref{fig:PofE}b). For abundance-weighted mean HLG, 99\% of the abiotic samples ranged between 11.55-11.75 eV (Supplementary Fig. \ref{fig:PofE}c) compared to the wider biotic range of 10.60-11.75 eV (99\%,  Supplementary Fig. \ref{fig:PofE}d). The abundance-weighted variance HLG showed the strongest separation, with abiotic samples remaining below 0.02 (99\% across all \textit{n} values,  Supplementary Fig. \ref{fig:PofE}e), while more than 50\% of the biotic samples exceeded 0.5 (for \textit{n} $\geq 7$,  Supplementary Fig. \ref{fig:PofE}f).

We then converted these likelihood distributions into confidence of biogenicity using Bayesian inference, requiring the specification of a prior probability P(B), which represents the expected likelihood of life before measuring the sample. To establish detection thresholds under different assumptions, we evaluated four prior probabilities ranging from equal (0.5) to highly skeptical (0.001). Using a prior of 0.5, Gini coefficients above 0.0075 yielded $\sim99\%$ confidence of biogenicity across all \textit{n} values (Supplementary Fig. \ref{fig:EqualPriorMetrics}b). Abundance-weighted mean values below 11.45 eV provided $\sim99\%$ confidence of biogenicity across all \textit{n} values, relaxing slightly to $\sim11.6$ eV at larger sample sizes (Supplementary Fig. \ref{fig:EqualPriorMetrics}d). Lastly, weighted variance values above 0.75 yielded $\sim99\%$ confidence (Supplementary Fig. \ref{fig:EqualPriorMetrics}f), with the threshold decreasing at higher sample sizes. Lower prior probabilities (0.1, 0.01, and 0.001) reduced the maximum achievable confidence (Fig. \ref{fig:LUMOSframework}b, Supplementary Fig. \ref{fig:EffectPrior}), particularly at small sample sizes (\textit{n} $\leq 7$). However, at a prior of 0.01, we still observe high confidence of biogenicity with values above 0.1 with sample sizes of \textit{n} $\geq 10$ (Fig. \ref{fig:LUMOSframework}b). At a lower prior (0.001), confidence transitions became abrupt, and while no single threshold achieves 99\% confidence across all sample sizes, specific regions of the parameter space still provide reliable biosignature detection (Supplementary Fig. \ref{fig:EffectPrior}f).

\section{Discussion}\label{Discussion}

The broader range of AAs HLG values observed in biotic samples points to an adaptive capability of biological systems to utilize a wider spectrum of energy gaps, regulating when and how electrons flow between molecular orbitals during reactions. Such diversity in electronic structure likely reflects the influence of complex metabolic networks, where molecules with varying reactivity coexist and participate in a myriad of interconnected biochemical processes (Supplementary Discussion \ref{supdis:contrastbiotic}). In contrast, abiotic chemistry is constrained by thermodynamic and kinetic limitations, producing AAs with a distinctive narrow range of HLG values that reflect limited chemical synthetic pathways and the inability to replenish molecules lost to degradation (Supplementary Discussion \ref{supdis:contrastabiotic}). 

We leveraged this distinction in HLG variability, quantified as weighted variance, within the LUMOS framework to determine the biotic or abiotic provenance of samples with high accuracy. Grounded in the quantum-mechanical nature of the HLG, the LUMOS framework is robust at classifying provenance of samples with $>$95\% accuracy, and adaptable to AA analyses from different extraction methods and analytical instruments (Fig. \ref{fig:LUMOSframework}c, Supplemental Dataset 1, Extraterrestrial\_Samples\_Info and Terrestrial\_Samples\_Info). Its non-instrument specificity indicates potential adaptability to future spaceflight instruments capable of molecule identification, including quantum tunneling-based detectors such as the Electronic Life-detection Instrument for Enceladus/Europa (ELIE), whose AAs detection patterns correlate with the HLG \cite{carr_solid-state_2023}. Importantly, because our method relies solely on the intrinsic electronic properties of molecules, it remains agnostic to a specific biochemistry and could be extended to evaluate the reactivity distributions of alien molecular systems, independent of Earth-centric assumptions about life's chemistry. This universality makes it relevant to proposed Ocean Worlds life-detection missions such as Enceladus Orbilander \cite{mackenzie_enceladus_2021}.

To further strengthen the robustness of LUMOS, it will be necessary to expand the characterization of abiotic environments and the molecular diversity of analyzed samples. Our current database remains biased toward a limited set of meteorite parent bodies \cite{broz_young_2024}, underscoring the need for a broader exploration of extraterrestrial environments and synthesis conditions. How these conditions may affect the HLG of AAs also remains largely unknown, yet characterizing this relationship will be critical for assessing the robustness of LUMOS across diverse planetary environments. Moreover, the preservation of the HLG signal across planetary environments is currently uncharacterized; understanding how degradation and planetary processes reshape HLG values over time will therefore be essential, and could be best achieved using Earth fossil positive controls as a training set, with planetary environments as the test set.  

LUMOS can be applied in the near term to life detection on Mars, where a putative biosignature was recently identified \cite{hurowitz_redox-driven_2025}. While experimental work has shown that amino acids can be rapidly degraded by irradiation, they may remain detectable at depths of 2 m or less for exposure ages up to $\sim$80 Myr \cite{pavlov_rapid_2022}, a representative exposure age of mudstone in Gale Crater \cite{farley_situ_2014}. The ExoMars Rosalind Franklin rover, expected to arrive on Mars in 2030, could reach this depth, and its Mars Organic Molecule Analyzer (MOMA) instrument has the ability to detect and quantify amino acids \cite{vago_exomars_2016}, enabling the application of LUMOS.

Confidence in LUMOS biogenicity assessments will further improve as the number of AAs targeted in both terrestrial and extraterrestrial sample analyses increases and non-detections are systematically reported, because each detected or absent molecule refines the weighted variance calculations. Integrating these experimental and observational insights with ab initio simulations of organic synthesis \cite{saitta_miller_2014} will help establish a comprehensive abiotic baseline, which will result in an step forward toward distinguishing biosignatures from background chemistry. This will provide a robust framework to assess the reactivity distributions of preserved amino acids as an agnostic and quantitative means of recognizing life beyond Earth.

\backmatter

\section{Methods}\label{Methods} 

\subsection{Amino acid database}\label{methods:database}
A comprehensive, literature-based database of amino acid abundances (nmol/g) for 56 amino acids extracted from diverse biotic and abiotic sources was compiled (Supplemental Dataset 1, Database). The dataset includes 102 abiotic samples, comprising samples from 14 meteorite groups \cite{aponte_analysis_2020,burton_propensity_2012,burton_extraterrestrial_2013,burton_effects_2014,burton_amino_2014,burton_amino_2015,callahan_search_2013,chan_amino_2012,elsila_amino_2021,glavin_amino_2006,glavin_effects_2010,glavin_extraterrestrial_2010,glavin_abundant_2020,glavin_extraterrestrial_2021,koga_extraterrestrial_2021,martins_indigenous_2007,martins_amino_2015,monroe_soluble_2011,simkus_low_2021}, lunar regolith \cite{elsila_origin_2016,elsila_soluble_2024}, asteroids \cite{parker_extraterrestrial_2023}, and two early-Earth laboratory simulations \cite{koga_extraterrestrial_2021}, alongside 87 environmental samples spanning 10 distinct biomes \cite{aerts_biosignature_2020,botta_polycyclic_2008,fuchida_concentrations_2015,hou_determination_2009,kobayashi_biomarkers_2022,noell_subcritical_2018,sane_amino_2020,takano_situ_2005,warren_variation_2017} and four biological kingdoms—Animalia \cite{abdul-hamid_nutritional_2002,caligiani_composition_2018,daniello_free_1995,hou_enzymatic_2011,hou_optimization_2011,kato_enantioselective_2015,liceaga-gesualdo_functional_1999,nilsang_optimization_2005,pascual-anaya_free_2006,tarbit_hippocampal_1980,wasswa_influence_2007,yoshikawa_free_2022}, Plantae \cite{ali_determination_2014,ares_determination_2022,origbemisoye_nutritional_2024}, Fungi \cite{mattila_basic_2002}, and Bacteria \cite{furlan_microalgae_2024}. The environmental samples represent biomes that include some of the closest terrestrial analogs to the Martian surface, as well as predicted subsurface environments of Enceladus and Europa. In addition, we compiled a second database comprising detections of 40 amino acids synthesized under a range of simulated conditions of environments throughout the Solar System's geologic history (e.g., interstellar clouds, Hadean Earth, Noachian Mars, Europa, and Titan) \cite{levy_prebiotic_2000,horst_formation_2012,hudson_amino_2008,johnson_miller_2008,kobayashi_formation_2023,neish_titans_2010,parker_primordial_2011,parker_plausible_2014,takeuchi_impact-induced_2020}. Since abundance data were unavailable for many of these studies in the second database, we recorded only the presence or absence of the amino acids reported in each work. This presence/absence dataset was then merged with the previously compiled database, excluding abundances, to yield a combined total of 233 samples and 64 amino acids. The first database was used for abundance-weighted analysis, while the second dataset was used for unweighted, presence-absence analysis. 

\subsection{Quantum-chemical calculations}\label{methods:qc}
The quantum-chemical properties of the amino acids were calculated using Psi4 (1.9.1 release) \cite{smith_psi4_2020}, and MOPAC (Molecular Orbital PACkage, 2016 release) \cite{stewart2016mopac}, the latter integrated within the Chem3D 23.1.1 software suite (PerkinElmer). The quantum-chemical properties calculated for purposes of this work included the energies of the HOMO and LUMO orbitals, and the HLG.

Chemical structures were initially constructed and optimized for geometry through energy minimization employing a modified Allinger's MM2 molecular mechanics algorithm, as implemented in Chem3D. Subsequently, quantum-chemical calculations of the electronic energy levels mentioned above were conducted using semi-empirical and ab initio (e.g. density functional theory) methods. The methods used for orbitals and electronic properties calculations included:

\begin{enumerate}[i.]
\item \textbf{Semiempirical Method:} The Modified Neglect of Diatomic Overlap (MNDO) supported by MOPAC 2016 within Chem3D 23.1.1 was employed. Calculations were conducted using the default software settings, including the closed-shell (restricted) method for ground-state calculations and the Eigenvector Following routine for geometry optimization. The Relative Maximum Amplitude (RMA) was set to 0.1, and the Mulliken method was chosen for calculating the distribution of electronic charge among the atoms. The COSMO solvation model was applied using water’s dielectric constant \(\epsilon = 78.4\).

\item \textbf{Density Functional Theory (DFT):} Single-point energy calculations were performed using the restricted closed-shell Hartree-Fock (RHF) wavefunction with the following combination of DFT functional and basis set: wB97XD/def2-TZVP. Additionally, the calculations were performed using the Polarizable Continuum Model (PCM) solvent model with water as the solvent.
\end{enumerate}

It's worth noting that a single conformer was optimized per amino acid, and enantiomers (L- and D-forms) were oprimized independently without enforcing mirror-related geometries. As a result, small differences in HLG values reflect convergence to different local conformational minima. In cases were chirality was not reported in a sample within the database, the average values of D- and L- forms of the amino acids were determined. 

\subsection{Additional Amino Acids Descriptors and Indexes}\label{methods:descriptors}
In addition to evaluating the HLG, a range (9 total) of physical and chemical molecular descriptors was assessed for all the amino acids within the database. Molecular energies, specifically the heats of formation and Gibbs free energies, were obtained through the ChemPropPro toolset in Chem3D 23.1.1. Molecular hydrophobicity (logP) and solubility (logS) values were also estimated using the ChemPropPro toolset. The dipole moment was obtained from the $\omega$B97XD/def2-TZVP/PCM(Water) Psi4 results and the topological polar surface area (TPSA) was obtained using the Mordred software \cite{moriwaki_mordred_2018}. Furthermore, the Molecular Assembly (MA) Index was calculated using the AssemblyGo tool \cite{jirasek_investigating_2024}. Finally, the carbon number and molar masses were derived from each amino acid's chemical formula.

\subsection{Class Separation Statistical Metrics Implementation}\label{methods:separation}

To assess the ability of each of the 10 molecular descriptors to separate between life and non-life, the symmetric relative entropy (Supplementary Methods \ref{supmethods:entropy}) was calculated between the abiotic (combined with simulated abiotic) and biotic distributions for each descriptor individually (unweighted-analysis). The pairwise comparison between the subcategories was calculated using MATLAB's \textit{ranksum} for Wilcoxon rank sum tests. We extended the unweighted-analysis analysis by incorporating the abundance (nmol/g) of each amino acid as extracted from the literature. For each descriptor, we weighted the values across all amino acids in a sample by their corresponding abundances. We used three statistical metrics to compute these weighted values: weighted mean (wmean) to capture the central tendency, weighted variance (wvar) to capture the distribution spread, and gini coefficient (gini) to capture inequality in the distribution. Finally, the samples were categorized as biotic and abiotic and the symmetric relative entropy between the categories was further calculated. Additional class separation methods evaluated included: (1) receiver operating characteristic (ROC) curve’s area under the curve (AUC) via MATLAB's \textit{perfcurve} function, (2) minimum redundancy maximum relevance (MRMR) ranking via MATLAB's \textit{fscmrmr} function, and (3) chi-square test univariate feature ranking using MATLAB's \textit{fscchi2} function.

\subsection{Machine Learning Classification}\label{methods:ml}

Machine Learning (ML) classification was performed using five representative models widely considered interpretable: fine, medium, coarse decision tree models, quadratic discriminant analysis, and binary generalized linear model logistic regression. These were implemented in MATLAB using the functions \textit{fitctree}, \textit{fitcdiscr}, and \textit{fitclinear}. We employed a cross validation strategy, repeated ten-fold; for each fold, the dataset was partitioned with a 20\% test set holdout and the remaining 80\% used for training. Sample class (biotic or abiotic) was used as the response variable. We then evaluated the performance of the five ML models using the weighted variance of the HOMO-LUMO gap (MNDO) as the sole predictor, and compared their performance to models trained on all 33 features in Fig. \ref{fig:performance}.

\subsection{Bayesian Simulation of Biogenicity Confidence}\label{methods:sml}

To estimate the confidence of biogenicity associated with abundance weighted variance, we generated biotic and abiotic populations representative of the empirical biotic and abiotic samples within our database. For each simulation, we selected \textit{n} amino acids (3-50) from those previously detected in samples of the corresponding provenance. The selection probability for each amino acid was scaled according to its occurrence frequency within the samples of the corresponding provenance. Abundance values were randomly generated within concentration ranges observed in the database for biotic and abiotic samples, preserving natural variability and analytical biases (e.g., mass spectrometry). For every simulated sample, we then computed an abundance-weighted variance based on its specific composition and abundance profile. After 1,000,000 simulation iterations, we estimated the probability density functions for both biotic and abiotic populations and applied Bayesian statistics to derive the likelihood of biogenicity given specific weighted variance values and amino acid counts. Details of the mathematical formulation are provided in Section \ref{supmethods:confidencebio}.

\backmatter

\section{Data, Materials, and Software Availability}

All data used in this study, including databases (amino acid profiles and properties), MATLAB scripts implementing our statistical analyses, and code for generating all figures presented in this article, are available at the following GitHub repository: https://github.com/jlramirezcolon/hlg-life-detection.

\section{Acknowledgments}

Funding was provided by NASA awards 80NSSC19K1028 and 80NSSC22K0188 to C.E.C. J.L.R.C. was supported by the NSF Graduate Research Fellowship Program. Z.N. acknowledges support from NASA award 80NSSC20K1092 to Lydia Bourouiba.

\bmhead{Author contributions} C.E.C and J.L.R.C. conceived of the strategy. J.L.R.C. compiled the database and conducted the quantum-chemical calculations. C.E.C. and J.L.R.C analyzed the data. Z.N. performed the simulations. All authors wrote the manuscript.

\bmhead{Competing interests}

The authors declare no competing interests.

\section{Supplementary Information}

\subsection{Supplementary Background}

\subsubsection{Abiotic Organic Synthesis} \label{ref:suppbackorganics}

A diverse array of organic compounds has been detected beyond Earth, spanning space and time. Nearly 150 molecules have been identified in the interstellar medium, with about a third containing six or more atoms \cite{herbst_complex_2009}. Laboratory examination of carbonaceous meteorites and asteroid-return samples has revealed a diverse suite of soluble organics in these extraterrestrial materials, including amino acids, carboxylic acids, amines, and nucleobases \cite{parker_extraterrestrial_2023, aponte_pahs_2023, glavin_abundant_2025}. These findings are further supported by the \textit{in situ} detection of organic molecules in various planetary bodies: thiophenic, aromatic, and long-chain organic compounds detected in ancient sedimentary rocks on Mars \cite{freissinet_long-chain_nodate,eigenbrode_organic_2018}, macromolecular organics captured from the active plume on Enceladus \cite{postberg_macromolecular_2018}, and carbon-nitrile species floating in the atmosphere of Titan \cite{waite_ion_2005}. Laboratory experiments have also demonstrated that many of these organic molecules can form abiotically through synthesis and/or degradation products thereof under plausible astrophysical and planetary conditions without life \cite{miller_production_1953, oba_nucleobase_2019}. Importantly, many of the abiotically produced organic compounds overlap structurally with those synthesized by biological systems. This overlap complicates the identification of definitive biosignatures, as many molecules traditionally associated with biology can also arise through abiotic pathways. 

\subsubsection{Previous Methods for Distinguishing the Origins of Organic Samples} \label{ref:suppbackapproaches}

Various approaches have been used to determine the biotic versus abiotic origin of a sample. One of the primary approaches is to analyze the isotopic composition and enantiomeric excess of molecules present in the sample. Life, as we know it, preferentially incorporates lighter isotopes and uses L-amino acids and D-sugars. Samples containing molecules with these isotopic characteristics and/or enrichment of these enantiomeric forms are likely indicative of biological origins. However, abiotic processes, such as Fischer-Tropsch reactions, can also produce molecules with similar isotopic signatures \cite{mccollom_carbon_2006}. Carbonaceous chondrites have also shown significant L-enantiomeric excesses in amino acids up to 60\% \cite{glavin_unusual_2012}, demonstrating that abiotic processes can produce chiral signatures. Furthermore, while strong and consistent enantiomeric excess across many amino acids could suggest a biological origin, such signatures have been shown to be susceptible to degradation and racemization on timescales $<$My in hydrothermally active oceans \cite{truong_decomposition_2019, steel_abiotic_2017} and in Martian brine analogs exposed to Martian conditions \cite{johnson_metal-catalyzed_2010}. These results imply that amino acids detected under these environments and conditions may be products of recent synthesis rather than preserved primordial material.

Another approach for distinguishing biotic and abiotic origins is to examine the distribution of amino acids present in a sample. Abiotic processes favor the synthesis of low mass amino acids with fewer carbon number due to the laws of thermodynamics \cite{dorn_monomer_2011}. In contrast, biotic processes synthesize and incorporate amino acids onto proteins based on their biological functions and thus show an abundance pattern that is independent of their carbon number. However, amino acid distributions can be altered over time due to preferential degradation of long-chain organic molecules, potentially resulting in a false negative pattern that resembles a distribution formed abiotically \cite{johnson_metal-catalyzed_2010}. One such approach for analyzing molecular frequency distributions involves machine-learning analysis of high-dimensional relationships among compound suites obtained through Pyr-GC-EI-MS spectra \cite{cleaves_robust_2023}. However, due to the high temperatures involved in pyrolysis, reaction products can rapidly recombine to form secondary compounds not originally present in the sample, limiting the method's ability to preserve a direct molecular signature of the source material. Beyond frequency distributions, the Molecular Assembly (MA) theory proposes that molecules requiring more assembly steps, defined by theoretical bond formation pathways, are less likely to arise from abiotic processes \cite{marshall_probabilistic_2017}. Therefore, by analysis of the MAIs of molecules within a sample, values are compared to a predicted threshold of 15, with values above this threshold indicating a likely biogenic origin \cite{marshall_identifying_2021}. However, its reliance on idealized valence rules and assembly pathways overlooks important factors, such as reaction conditions, kinetics, and thermodynamic constraints. As analytical techniques and space missions advance, new insights into both biotic and abiotic processes will continue to emerge, underscoring the need for robust, universal frameworks that can reliably distinguish between biotic and abiotic origins across diverse environments.

\subsection{Supplementary Methods}

\subsubsection{Symmetric Relative Entropy as a Measure of Class Separation} \label{supmethods:entropy} 

As a primary measure of class separation we used the ``divergence" between two probability measures \cite{kullback_leibler_1951}, now commonly referred to as the Jeffrey's Divergence or symmetric Kullback-Leibler (KL) divergence. For two distributions P and Q this is given by $D_J(P, Q) = D_J(Q, P) = D_{\text{KL}}(P \| Q) + D_{\text{KL}}(Q \| P)$, where $D_{\text{KL}}(P \| Q)$ is the KL Divergence of $P$ from $Q$, given by: 

\[D_{KL}(P \parallel Q) = \sum_{x \in X} P(x) \log\left(\frac{P(x)}{Q(x)}\right)\]

We refer to the Jeffreys divergence $D_J$ as symmetric relative entropy to avoid confusion with the non-symmetric KL-divergence commonly referred to as relative entropy. If we take P to be the distribution of a feature derived from the “non-life” or abiotic class, $D_{\text{KL}}(P,Q)$ represents the expected surprise from using values derived from the “life” or biotic class as a model (Q) instead of the “non-life” (abiotic) class. This provides a measure of how informative a given feature is in distinguishing biotic from abiotic. Likewise, $D_{\text{KL}}(Q,P)$ provides a measure of how informative a given feature is in distinguishing abiotic from biotic. A-priori, we don't know whether a given feature will have higher $D_{\text{KL}}(P,Q)$ or $D_{\text{KL}}(Q,P)$ so the Jeffrey's divergence is a reasonable starting point.

We estimated $D_J$ using MATLAB's \textit{relativeEntropy} function, which reports $D_J$ in nats. MATLAB utilizes a closed-form solution computed directly from samples, making this an approximate measure due to potential violation of the Gaussian assumption. However, this approach avoids the many problems of computing the Jeffrey's divergence from probability density functions (PDFs) on a common support (e.g., sparse data at extrema due to the small size of the dataset, non-overlapping or mismatched support, resolution and binning sensitivity, infinite or undefined values, and arbitrary nature of fixes such as smoothing, regularization, and truncation). The distributional assumption makes $D_J$ well defined for continuous-valued data with limited sample numbers and is a reasonable first-order assumption often used in feature selection, one we have also adopted. This implies that the estimated $D_J$ values may differ between different calculation methods. The important aspect here is the relative values of the $D_J$ estimates for different features, where higher values indicate greater separation between biotic and abiotic classes.

\subsubsection{Estimating the Confidence of Biogenicity} 
\label{supmethods:confidencebio}

We consider a life detection simulation where B is biogenicity, and A is abiogenicity, equal to the logical negation of B (¬B). We use a Bayes theorem formulation to permit incorporation of different priors. P(B) is the prior probability of biogenicity before incorporating new evidence, and P(B) = 1-P(A). For example, P(B) = 0.5 represents a neutral prior where life and non-life are considered equally likely. We also consider a lower prior because to detect life, an environment would need to be habitable, inhabited, preserve biosignatures, and they would need to be detectable based on a specific sample and analysis technology. Therefore, we also analyze P(B) = 0.01, 0.001, and 0.001. The following definitions apply:

\begin{itemize}
    \item $P(E)$ is the marginal likelihood, the probability of observing evidence E, regardless of whether life exists or not.
\end{itemize}

\begin{itemize}
    \item $P(E\mid B))$ is the likelihood of observing evidence E given biogenicity (B).
\end{itemize}

\begin{itemize}
    \item $P(E \mid A))$ is the likelihood of observing evidence E given abiogenicity (A).
\end{itemize}

Evidence E might arise from either an abiotic or biotic sample. By the rule of total probability, this is: 

\[
P(E) = P(E \mid B) \cdot P(B) + P(E \mid A) \cdot (1 - P(B))
\]

This result says that the probability of observing evidence E is related to our prior ${P}({B})$ and to the conditional probabilities ${P}({E}\mid{B})$ and ${P}({E}\mid{A})$. We want to estimate ${P}({E}\mid{B})$ and ${P}({E}\mid{A})$ under conditions of imperfect measurement, where we may only be able to measure k amino acids. Here we select the k amino acids at random, and it is left as an exercise for the future to evaluate specific technologies and measurement approaches.

To compute ${P}({E}\mid{A})$, we consider only abiotic (A) samples in the database and select k AAs previously found in abiotic samples, with selection probability weighted by each amino acid's occurrence frequency among abiotic samples. Next, we assign each AA an abundance using a uniform distribution over its observed abundance range across all abiotic samples. We then calculate the feature of interest, e.g., the HLG weighted variance, and repeat the process 1,000,000 times. The resulting histogram is normalized to get a probability density function (PDF), our estimate of ${P}({E}\mid{A})$. We then repeat this process for the biotic (B) subset to estimate ${P}({E}\mid{B})$.

The probability of (e.g., confidence in) biogenicity B given evidence E can now be estimated using Bayes theorem:
\[
P(B \mid E) = \frac{P(E \mid B) \cdot P(B)}{P(E)}
\]
and after substituting our expression for ${P}({E})$ above,
\[
P(B \mid E) = \frac{P(E \mid B) \cdot P(B)}{P(E \mid B) \cdot P(B) + P(E \mid A) \cdot (1 - P(B))}
\]
and finally simplifying we have: 
\[
P(B \mid E) = \frac{P(E \mid B)}{P(E \mid B) + P(E \mid A) \cdot \left( \frac{1}{P(B)} - 1 \right)}
\]

\subsection{Supplementary Results}

\setcounter{figure}{0}
\renewcommand{\thefigure}{S\arabic{figure}}

\begin{figure}[H]
    \centering
    \includegraphics[width=1\textwidth]{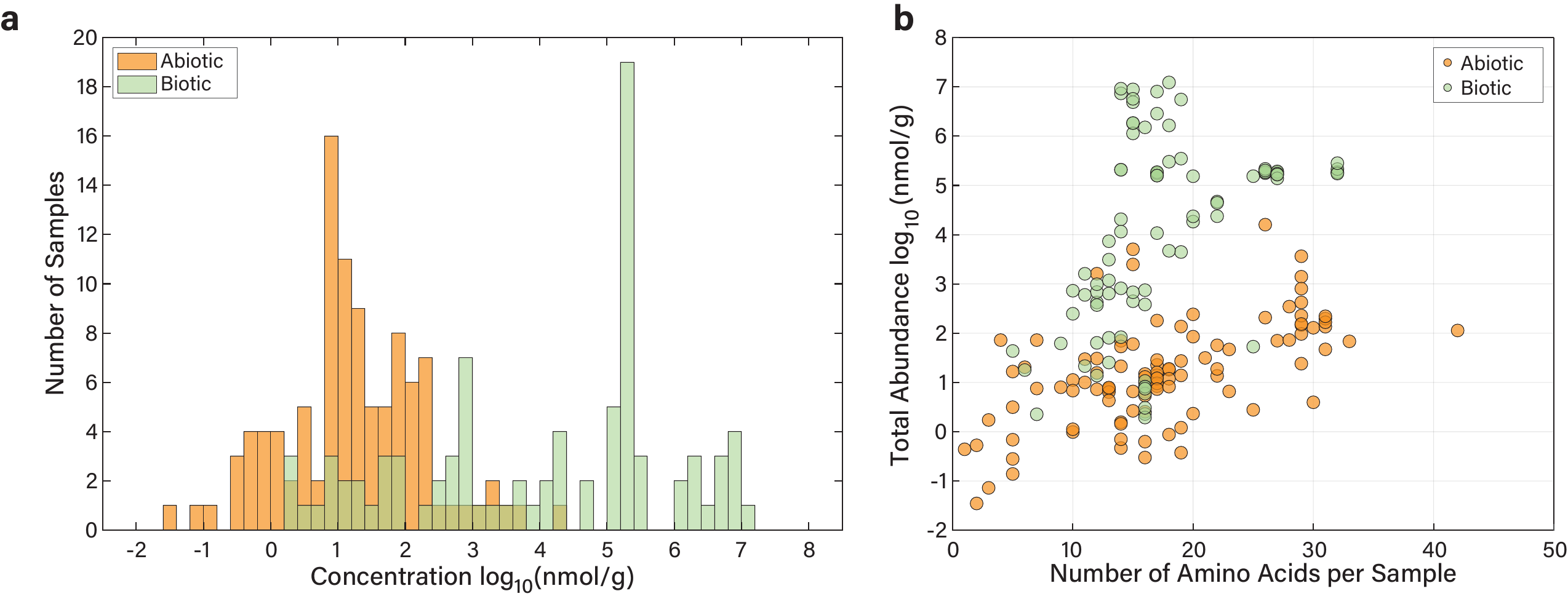}
    \caption{\textbf{Abundance of Amino Acids Across Biotic and Abiotic Classes.} \textbf{a}, Total concentration of amino acids (nmol$/$g) per sample across biotic and abiotic classes. \textbf{b}, Relationship between the number of amino acids detected and their total abundance (nmol$/$g) per sample in biotic and abiotic classes.}
    \label{fig:TotalAbundance}
\end{figure}

\begin{figure}[H]
    \centering
    \includegraphics[width=1\textwidth]{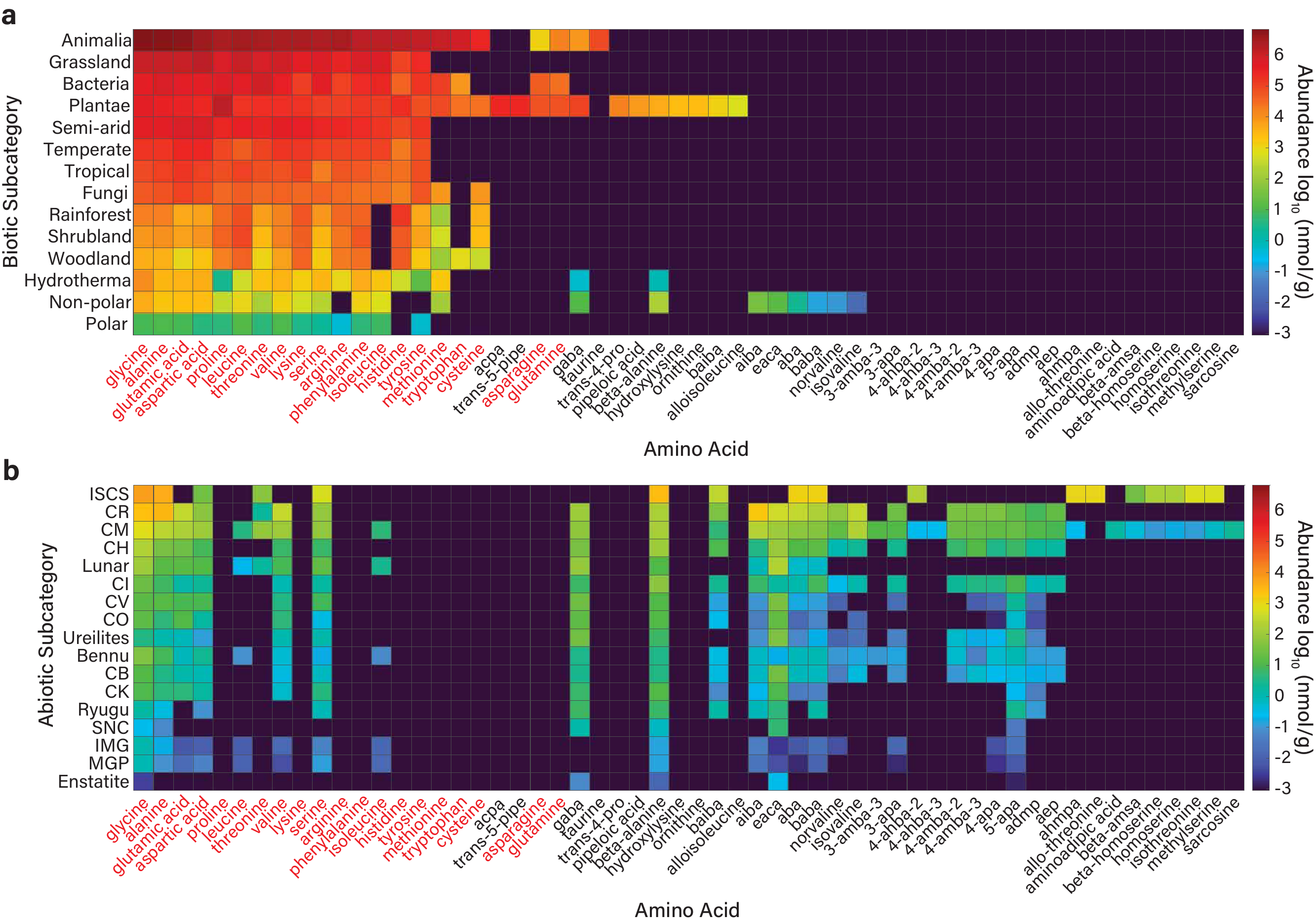}
    \caption{\textbf{Amino Acid Abundance Patterns by Biotic and Abiotic Subcategories.} \textbf{a}, Total amino acid abundances (nmol$/$g) per biotic subcategory. Amino acids are ordered from most to least abundant (left to right). Subcategories are arranged from highest to lowest total abundance (top to bottom). Proteinogenic amino acids are highlighted in red. \textbf{b}, Total amino acid abundances (nmol$/$g) per abiotic subcategory, with amino acids plotted in the same order as in (a) for direct comparison and subcategories plotted in the same format as (a). Refer to Supplemental Dataset 1 (Database) for full amino acid names and subcategory names.}
    \label{fig:AbundanceHeatmap}
\end{figure}

\begin{figure}[H]
    \centering
    \includegraphics[width=1\textwidth]{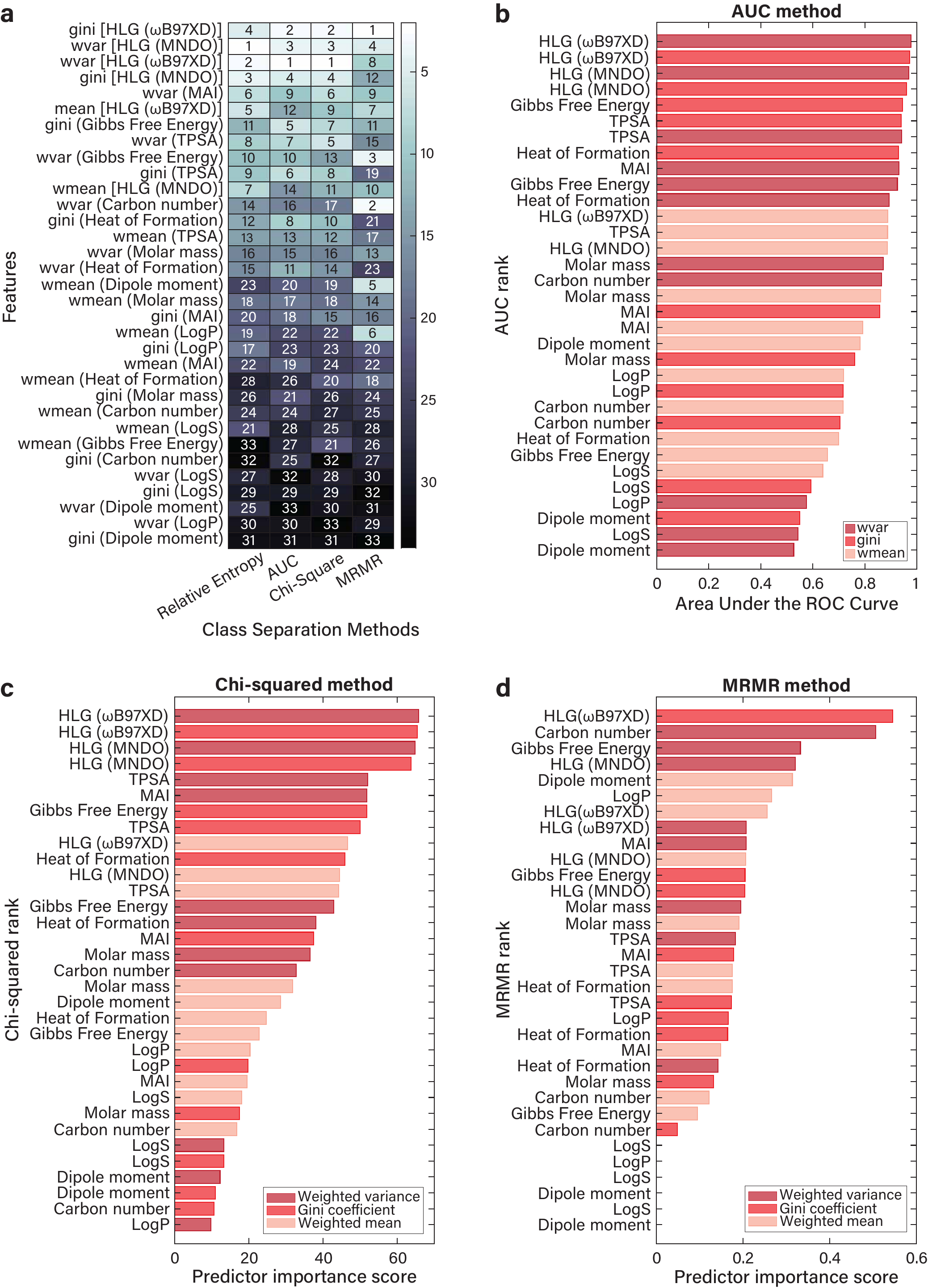}
    \caption{\textbf{Evaluation of Alternative Class Separation Methods Using Abundance-Weighted Amino Acid Descriptors to Differentiate Biotic and Abiotic Samples.} a, Overall performance ranking of molecular descriptors (by weight-based statistical metrics) based on the average of their individual rankings across all class separation methods. Alternative class separation methods included: \textbf{b}, Area Under the Receiver Operating Characteristic Curve (AUC); \textbf{c}, Chi-squared; and \textbf{d}, Minimum Redundancy Maximum Relevance (MRMR).}
    \label{fig:ClassificationMethods}
\end{figure}

\begin{table}[h]
\caption{Test accuracies (\%) of model types by selected feature(s).}\label{tab1}%
\begin{tabular}{@{}llll@{}}
\toprule
Model & wvar (HLG-MNDO) & All features \\
\midrule
Fine Tree & 96.84 ± 2.42 & 95.26 ± 2.08 \\
Medium Tree & 96.84 ± 2.42 & 96.84 ± 2.42 \\
Coarse Tree & 96.84 ± 2.42 & 96.84 ± 2.42 \\
Quadratic Discriminant & 96.05 ± 3.77 & 95.79 ± 3.55 \\
Binary GLM Logistic Regression & 85.79 ± 5.44 & 61.58 ± 5.98 \\
\botrule
\end{tabular}
\label{table:testaccuracies}
\end{table}

\begin{figure}[H]
    \centering
    \includegraphics[width=1\textwidth]{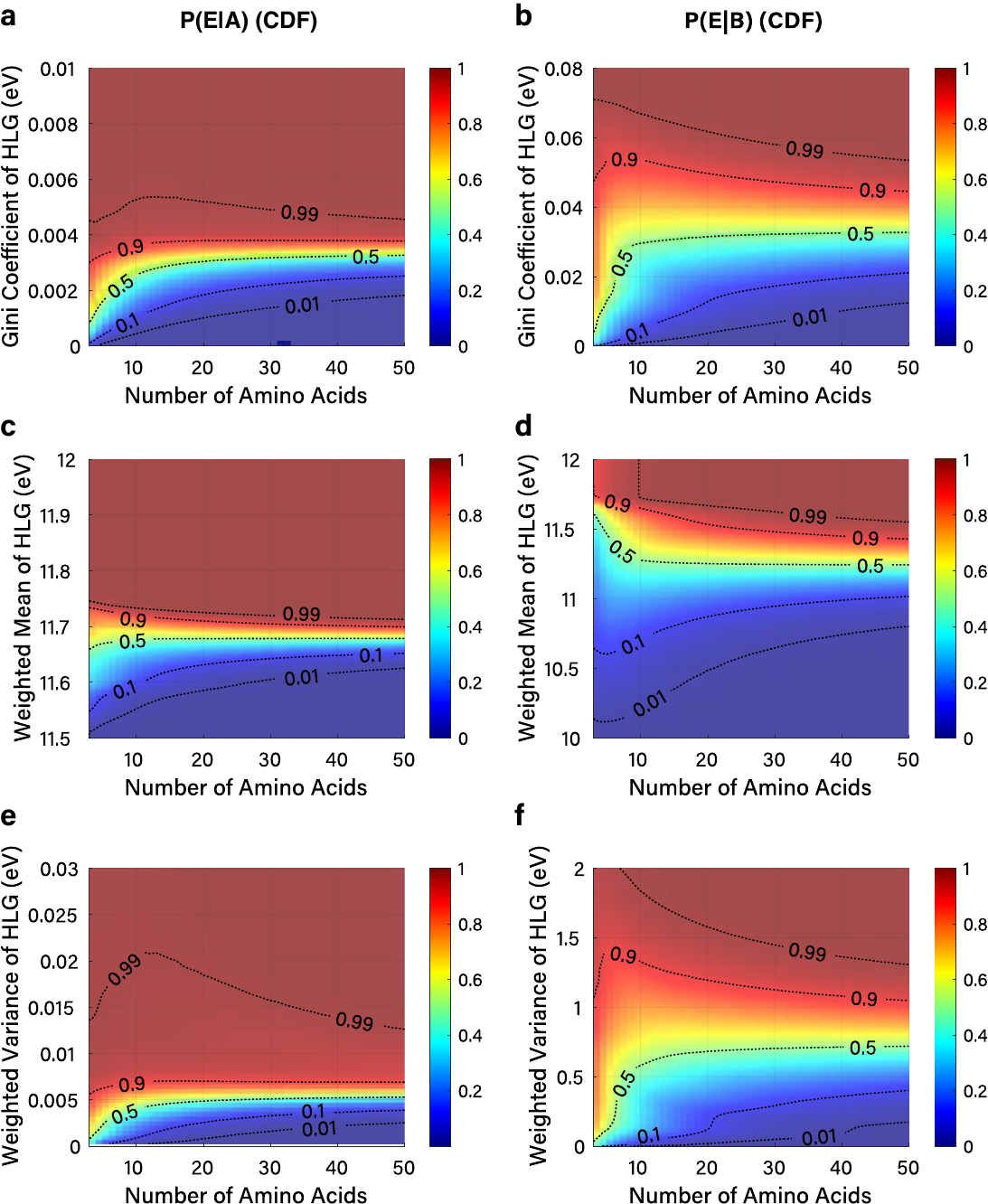}
    \caption{\textbf{Probability of evidence given abiotic and biotic datasets}. Heatmaps show the cumulative probability of observing evidence (E) given abiogencity \textit{P(E$\mid$A)} (\textbf{a}, \textbf{c}, and \textbf{e}), or biogenicity \textit{P(E$\mid$B)} (\textbf{b}, \textbf{d}, \textbf{f}) based on Monte carlo simulations (\textit{n} = 1,000,000) using compiled abiotic and biotic amino acid sample databases.  Panels show likelihoods calculated from (\textbf{a}, \textbf{b}) Gini coefficient, (\textbf{c}, \textbf{d}) abundance-weighted mean, and (\textbf{e}, \textbf{f}) abundance-weighted variance of the HLG (MNDO). The x-axis represents the number of amino acid species detected in a  simulated sample, while the y-axis represents the obtained metric value for that sample. Color intensity reflects the probability of observing the metric value given the number of amino acids, with iso-probability contours overlaid.}
    \label{fig:PofE}
\end{figure}

\begin{figure}[H]
    \centering
    \includegraphics[width=1\textwidth]{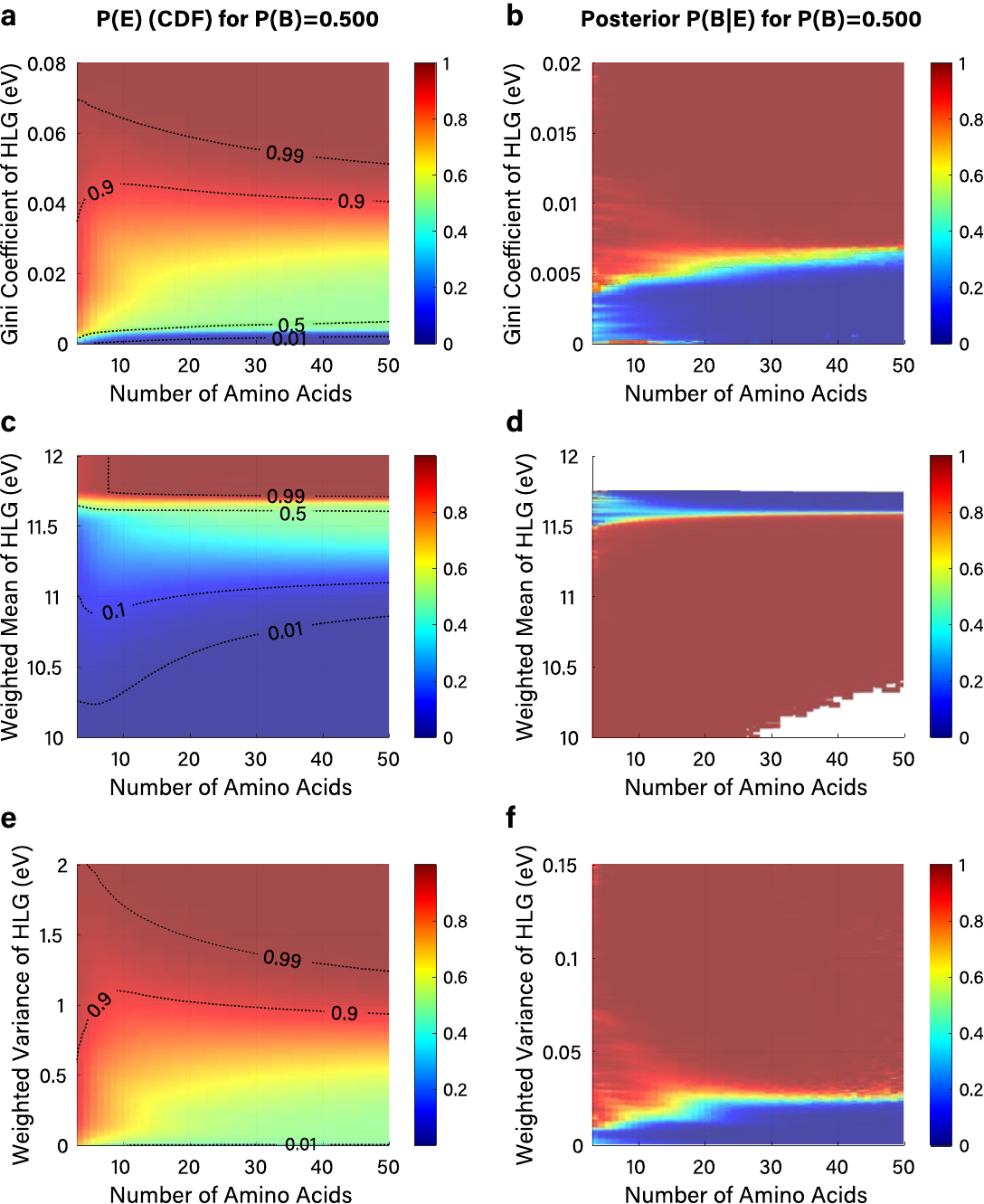}
    \caption{\textbf{Confidence of biogenicity at prior of 0.5 across metrics}. Left panels show the cumulative probability P(E) for observing metric values across all samples (abiotic or biotic) for (\textbf{a}) Gini coefficient, (\textbf{c}) weighted mean, and (\textbf{e}) weighted variance. Right panels show the probability that a sample is biotic given its observed metric value, calculated assuming equal prior likelihood of abiotic versus biotic origin (0.5) for (\textbf{b}) Gini coefficient, (\textbf{d}) weighted mean, and (\textbf{f}) weighted variance. The x-axis represents the number of of amino acids species detected. Color intensity in the right panels indicates confidence in biogenicity, with warmer colors representing higher probability that the sample is biotic of origin.}
    \label{fig:EqualPriorMetrics}
\end{figure}

\begin{figure}[H]
    \centering
    \includegraphics[width=1\textwidth]{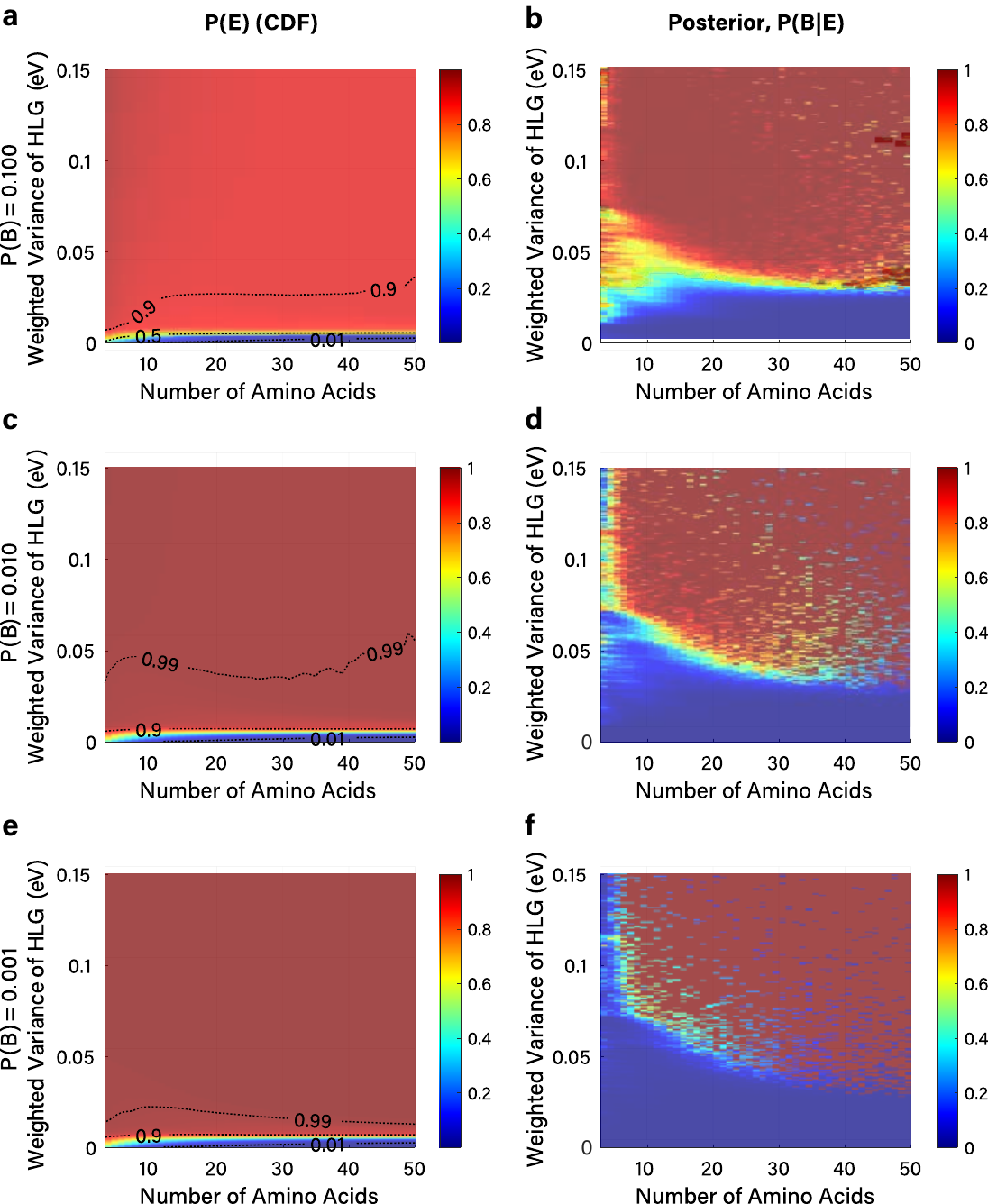}
    \caption{\textbf{Effect of prior on evidence distribution and posterior confidence in biogenicity}. Left panels show the cumulative probability of observing evidence \textit{P(E)} based on abundance-weighted variance of HLG for prior probabilities of biogenicity \textit{P(B)} of  (\textbf{a}) 0.1,  (\textbf{c}) 0.01, and  (\textbf{e}) 0.001. Right panels show the corresponding posterior probabilities P(E$\mid$B) for  (\textbf{b}) 0.1,  (\textbf{d}) 0.01, and  (\textbf{f}) 0.001.}
    \label{fig:EffectPrior}
\end{figure}

\subsection{Supplementary Discussion}

\subsubsection{Weighted variance strengths and limitations}\label{supdis:wvar}

The superior class separation power of weighted variance was particularly evident for HLG, MAI, TPSA, and Gibbs Free energy, which were also the four top-performing descriptors overall. This result highlights that variability in these physicochemical properties differs more consistently between biotic and abiotic amino acid samples than their average values or inequality (Gini coefficient) distributions. However, this metric faces two key statistical limitations as shown by the false negative samples in our analysis. First, weighted variance is sensitive to abundance skew, where a few dominant compounds can obscure signals from lower-abundance species. Second, the metric requires adequate molecular diversity for robust characterization. Samples with fewer than 10 amino acid species may lack the statistical power to reliably distinguish biotic from abiotic patterns. 

Analysis of the five overlapping samples (Fig. \ref{fig:performance}b) reveals how these limitations operate in practice. One case, from Antarctic ice \cite{botta_polycyclic_2008}, was associated with a collected meteorite and the most abundant measured amino acid was $\alpha$-aminoisobutyric acid ($\alpha$AIB), a non-proteinogenic amino acid commonly found in extraterrestrial samples. This sample therefore represents a likely class label error rather than a putative false negative. Two other samples were from the Atacama Desert \cite{moore_atmospheric_2024}, specifically the sites North of Antofagasta (NOA) and Diego de Almagro (DGA). The NOA sample had no measurable glycine, and the highest abundance amino acid was $\epsilon$-aminocaproic acid, with a high HLG (11.5 eV), a known contaminant derived from nylon-6 \cite{glavin_amino_2006}; the second highest abundance was $\beta$-alanine, which could be a degradation product or input from meteoritic sources; two non-proteinogenic amino acids, $\gamma$-aminobutyric acid (GABA) and AIB, were also detected in this sample. This sample is interpreted as representing plastic contamination combined with breakdown products and possible meteoritic input. The DGA sample was dominated by glycine and $\epsilon$-aminocaproic acid ($\epsilon$-aca), along with other proteinogenic amino acids. This sample is interpreted as representing plastic contamination overprinting biotic-origin amino acids. Finally, two samples from mouse macrophages appear as false negatives, since only six amino acids (all with relatively high HLGs) were measured. 

When contamination is minimized and an informative range of amino acids is included, as is the case for the majority of samples in our dataset, HLG-weighted variance may achieve accuracy up to 100\%. These findings emphasize that the pattern of variation in molecular properties, represents a robust statistical feature distinguishing biotic from abiotic chemistry, provided that samples contain sufficient molecular diversity and lack extremely skewed abundance distributions.

\subsubsection{Performance of other molecular descriptors}\label{subdis:descriptors}

\textbf{\textit{Molecular Assembly Index (MAI)}}. Our class separation analysis identified MAI as a descriptor with moderate-to-strong class separation capability between the biotic and abiotic categories in the unweighted and weighted analyses (Supplemental Dataset 1, Statistics\_Tables, Table 6). The MAI quantifies molecular complexity and information content by measuring the number of steps or combinations required to build a molecule from smaller units \cite{marshall_probabilistic_2017}. Although \citet{marshall_identifying_2021} proposed a MAI threshold of 15 to distinguish abiotic from biotic origins, citing a probability of one in 10²³ for synthesis of a molecule with a MAI $\geq 15$, all amino acids in our dataset exhibited lower values, ranging from 3 to 11 in the biotic category and 3 to 9 in the combined abiotic category. Consequently, the MAI performed more strongly in the abundance-weighted analysis, where biotic samples contained more abundant high-MAI amino acids and abiotic samples favored lower-MAI amino acids. These results highlight the limitations of the MAI as a standalone classifier when restricted to relatively simple organic molecules common to both origins. Nonetheless, we argue that, like the HLG, the MAI captures functionally relevant complexity that may reflect evolutionary selection pressures acting on biotic environments. 

\textbf{\textit{Topological Polar Surface Area (TPSA)}}. Similar to MAI, TPSA was identified as a descriptor with strong-to-moderate class separation performance in the abundance-weighted analysis, but poor separation capability in the unweighted analysis. TPSA quantifies the surface occupied by polar atoms, primarily nitrogen and oxygen. The unweighted analysis revealed that both abiotic and biotic samples contained amino acids spanning a similar TPSA range (60-30 \text{\AA}$^2$), reflecting varying degrees of polarity. However, the weighted-TPSA features, particularly the weighted mean, showed that abiotic samples tend to cluster near the boundary between hydrophobic and moderately polar regions while biotic samples are skewed toward higher polarity. This is further supported by the Gini coefficient-TPSA feature, which indicates a more uniform TPSA distribution in abiotic samples (low Gini values) and a more polarized distribution in biotic samples, likely driven by higher abundances of polar amino acids. Despite these trends, the MRMR analysis suggests that TPSA contributes little unique information beyond what is already captured by the HLG. Nevertheless, we posit that TPSA may become increasingly informative when applied to larger, more functionally diverse molecules such as peptides.

\textbf{\textit{Gibbs free energy and Heat of formation}}. Gibbs free energy quantifies the maximum usable energy available to perform work in a system under constant temperature and pressure, incorporating both enthalpy and entropy contributions. The enthalpy component, measured by the heat of formation, reflects the energy change associated with forming one mole of a compound from its constituent elements in their standard states. In the unweighted and weighted class separation analysis, both thermodynamic parameters exhibited moderate separation capability between biotic and abiotic samples, with significant overlap between the two classes. These findings suggest that both biotic and abiotic environments favor amino acids within a similar thermodynamic stability range, likely reflecting shared chemical constraints. Notably, only biotic systems in our dataset include amino acids with positive Gibbs free energy values, thus highlighting the capacity of life to overcome thermodynamic barriers through the use of energy-coupling mechanisms, such as ATP hydrolysis.

\textbf{\textit{Carbon number and Molar mass}}. These two molecular descriptors demonstrated moderate-to-low performance in separation between biotic and abiotic samples, with notable differences in the unweighted and abundance-weighted analyses. Carbon number ranked as the second-best descriptor unweighted analysis (Fig. \ref{fig:HLGdistribution}d), yet its performance dropped significantly when abundance was incorporated. Weighting metrics such as the weighted mean and Gini coefficient revealed small separation between classes, though weighted variance did suggest distinct distributions, with biotic samples displaying greater variability in carbon number and molar mass (Supplementary Figure \ref{fig:ClassificationMethods}a). 

Despite moderate-to-low performance in most of the class separation methods, carbon number ranked as the second most informative feature in the MRMR analysis (Supplementary Figure \ref{fig:ClassificationMethods}d). This indicates that it adds incremental predictive value beyond what the HLG feature provides, which ranked first. This likely reflects its ability to represent molecular scale and structural complexity, such as atom and chain length, which are only indirectly reflected in the HLG. Overall, these size-based descriptors offer limited classification capability on their own, but may contribute when combined with more functionally or electronically informative descriptors. 

\textbf{\textit{LogP and LogS}}. Both parameters exhibited poor class separation performance across the unweighted and weighted analyses. LogP, which estimates hydrophobicity based on the octanol-water partition coefficient, showed similar distributions between abiotic and biotic categories, though abiotic samples spanned a broader range. Despite its poor individual performance, LogP ranked sixth in the MRMR analysis (Supplementary Figure \ref{fig:ClassificationMethods}d), suggesting it may have contributed minimal, but complementary information when combined with other features. LogS, a predicted solubility metric derived from structural and polarity features, performed similarly poorly. Its value distributions were highly overlapping between classes, and no consistent trend emerged across weighted analyses.

\textbf{\textit{Dipole moment}}. This descriptor demonstrated low class separation performance in the unweighted and weighted analyses, ranking as the worst-performing descriptor across all class separation methods. Defined as a measure of the separation of positive and negative charge within a molecule, dipole moment reflects overall molecular polarity and is typically influenced by electronegativity differences and molecular geometry. The unweighted distributions were nearly indistinguishable across biotic and abiotic categories, and while a slight rightward shift in the weighted mean was observed for biotic samples, this trend appeared to be driven by a small number of high-value outliers (e.g., taurine and histidine). Notably, dipole moment ranked fifth in the MRMR analysis (Supplementary Figure \ref{fig:ClassificationMethods}d), which may reflect its disproportionate influence in specific samples (e.g., high-value outliers) rather than a consistent class-level trend.

\subsubsection{Comparison with previous studies assessing HLG distributions} 

A recent study by \citet{abrosimov_homo-lumo_2024} compared HLG values between 134 abiotic organic molecules from the Murchison meteorite and 570 microbial and plant secondary metabolites. Their analysis revealed that biotic molecules exhibited significantly narrower and smaller HLG values (10.4 ± 0.9 eV) compared to abiotic molecules (12.4 ± 1.6 eV). Our results show similar average values (within standard deviations) for the biotic set (10.68 ± 0.51 eV) but significantly lower average values for the abiotic set (10.85 ± 0.27 eV). However, direct comparisons are limited because of the use of different computational methods (AM1 versus wB97XD/def2-TZVP) for determining HLG values and our focus being solely on AAs. Moving forward, standardizing computational approaches using high-accuracy quantum chemical methods will be essential not only for enabling meaningful comparison across an expanding set of biotic and abiotic samples but also for facilitating collaborative research efforts and improving the use of the HLG distributions as a biosignature.

\subsubsection{Biotic innovation in amino acid electronic properties}\label{supdis:contrastbiotic}

Biological systems use amino acids across diverse molecular environments where enzymatic machinery can stabilize and direct reactions involving chemically reactive species. This cellular control allows organisms to employ amino acids with electronic properties that would be instable in abiotic settings. For example, aromatic systems in tryptophan facilitate $\pi$-stacking interactions essential for protein stability \cite{riley_importance_2013}; histidine's imidazole group group enables metal binding and catalysis \cite{dokmanic_metals_2008}; cysteine's sulfyhydryl group contributes to redox sensing \cite{wang_redox_2012,jones_mapping_2011}; and arginine's guanidino group mediates substrate recognition \cite{vagenende_protein-associated_2013,luscombe_amino_2001}. The broader HLG distribution observed in biotic samples is consistent with \citeauthor{granold_modern_2018}'s \citeyear{granold_modern_2018} hypothesis that protein-building amino acids evolved to become systemically ``softer" (lower HLG) and more redox-reactive \cite{granold_modern_2018}. This suggests that organisms have optimized electronic properties to achieve the balance between stability and reactivity required for complex biochemical processes.

\subsubsection{Abiotic constraints in amino acid electronic properties}\label{supdis:contrastabiotic}

We found that abiotically-formed amino acids exhibited a relatively constrained range of HLG values, with an observed lower boundary of approximately 10 eV. We propose that this boundary likely represents a fundamental stability threshold governing the persistence of amino acids in abiotic environments. Amino acids with lower HLG values would be more reactive and susceptible to degradation due to their decreased kinetic stability across the range of conditions represented in our database (Supplemental Dataset 1, Database). While abiotic pathways could theoretically produce most proteinogenic amino acids- including 17 of 20 via the Strecker cyanohydrin synthesis pathway \cite{chimiak_carbon_2021, lerner_iminodicarboxylic_2005, lerner_strecker_1993} given specific aldehyde precursors \cite{cobb_natures_2015} and aromatic proteinogenic amino acids via water-rock catalyzed Friedel-Crafts alkylation \cite{menez_abiotic_2018}- only 15 have been detected in asteroid samples analyzed to date \cite{glavin_abundant_2025, parker_extraterrestrial_2023, potiszil_insights_2023, mojarro_prebiotic_2025}. The absence of the remaining amino acids like cysteine, glutamine, histidine, arginine, and lysine can thus be explained by either the absence of the aforementioned aldehyde precursors or their increased reactivity (lower HLG values) that prevents their accumulation and persistence. Furthermore, while tryptophan, tyrosine, and methionine (all low-HLG AAs) were detected in Bennu, the reported trace-level abundances support the prediction that high reactivity of such AAs limits their accumulation in abiotic environments.

\subsubsection{Instrument Applications with HLG} 

Mass spectrometry (MS) and microcapillary electrophoresis ($\mu$CE), combined with standards and controlled fragmentation, enable chemical formula assignment and polymer sequence determination. Both techniques have been developed for life detection applications \cite{chou_planetary_2021, duca_quantitative_2022, seaton_robust_2023}, with MS additionally employed for analysis of organics in extraterrestrial samples \cite{glavin_abundant_2020, glavin_abundant_2025, aponte_analysis_2020, parker_extraterrestrial_2023}. Advanced MS methods can further utilize transients to determine molecular collision cross sections \cite{james_advancing_2022}, better constraining molecular structures from which the HLG can be determined. Notably, nanogap tunneling current measurements of single molecules have demonstrated strong correlations with the HLG, at large gap sizes \cite{carr_solid-state_2023}. This technique enables detection of individual biomolecules, from amino acids to peptides and nucleic acid sequences, and could extend to agnostic detection, including polymers from unknown biochemistries. With nanogap-based instruments currently being developed for spaceflight applications, the direct measurement of electronic properties in extraterrestrial environments is becoming increasingly feasible, positioning HLG as a powerful descriptor within our LUMOS framework for sensitive and specific biotic and abiotic class separation. \cite{carr_solid-state_2023}.

\subsubsection{Risk of False Negatives in Life Detection}\label{supdis:fn}

The overlapping cases discussed in \ref{supdis:wvar} translate into mission-critical detection risks for life detection, where false negatives may represent the most consequential error type. The macrophage and Atacama samples reveal three specific failure scenarios. First, limited molecular diversity in samples with few detectable species reduces statistical power, a concern for ancient, degraded, or biomass-limited samples. Second, contamination obscuring biological signals can shift distributions toward abiotic-like patterns, whether from spacecraft materials or sample handling. Third, beyond the specific cases in our dataset, extraterrestrial agnostic life with different building block repertoires and unusual functional groups, might produce electronic property distributions that fall outside the biotic classification space defined in this study.

Reducing these risks requires combining multiple independent measurements beyond HLG (e.g., chirality, isotopic ratios, molecular complexity) along with strict contamination controls and detection approaches designed to minimize false negative. Rather than relying on any single metric, life detection mission will need to build cases for multiple techniques that together provide complementary lines of evidence.

\bibliography{sn-bibliography}

\end{document}